\documentclass[11pt,a4paper]{article}
\usepackage[utf8]{inputenc}
\usepackage{amsmath,amssymb,amsfonts}
\usepackage{graphicx}
\usepackage{geometry}
\usepackage{booktabs}
\usepackage{microtype}
\usepackage{hyperref}

\usepackage[nosort,compress]{cite} 

\title{\textbf{Longitudinal Development Analysis of Extensive Air Showers Using CORSIKA Simulations}}
\author{\textbf{Ali A. Shihab}$^{1,2}$ and \textbf{Al-Rubaiee A. A.}$^{2}$ \\
\small $^{1}$Ministry of Agriculture, Directorate of Animal Resource, \\
\small Department of Protection of Genetic Resource, Baghdad, Iraq.\\
\small $^{2}$Mustansiriyah University, College of Science, Department of Physics, Baghdad, Iraq.}
\date{}

\begin{document}

\maketitle

\begin{abstract}
We present a comprehensive analysis of the longitudinal development of Extensive Air Showers (EAS) simulated with CORSIKA version 7.7500 for proton, helium, and iron primaries at energies of $10^{15}$, $10^{16}$, and $10^{17}$~eV across zenith angles of $0^\circ$, $30^\circ$, and $45^\circ$. For each combination of primary type, energy, and zenith angle, 50 independent showers were simulated, resulting in a total of 450 simulated showers. The Gaisser–Hillas function was fitted to extract the depth of shower maximum ($X_{\max}$) and the number of particles at maximum ($N_{\max}$). Our results confirm the expected logarithmic increase of $X_{\max}$ with energy ($\sim 100$~g/cm$^2$ per decade), as well as systematically shallower $X_{\max}$ for heavier primaries (iron vs. proton: $\Delta X_{\max} \approx 160$~g/cm$^2$ at $10^{17}$~eV). The simulations also reproduce the expected $\sec(\theta)$ scaling behavior. Multi-component analysis reveals distinct evolutionary patterns for electromagnetic, muonic, and hadronic components. These findings provide benchmark-level simulations for cosmic-ray composition studies and validate the CORSIKA framework for multi-parameter analyses of air-shower development.\\

\end{abstract}

\section{Introduction}
When a high-energy cosmic-ray particle collides with a nucleus in the upper atmosphere, it initiates a cascade of secondary particles known as an Extensive Air Shower (EAS) \cite{Gaisser2013}. The development of this shower along its axis of propagation—the longitudinal profile—encodes information about the identity (mass number $A$), energy ($E$), and zenith angle ($\theta$) of the primary particle \cite{Dova2003}. The two crucial parameters are the depth of the shower maximum ($X_{\max}$) and the number of particles at maximum ($N_{\max}$), both expressed as functions of atmospheric depth.

The atmospheric depth $X$ traversed by a shower incident at a zenith angle $\theta$ is given by:
\begin{equation}
X(h,\theta) = \frac{X_{\text{vertical}}(h)}{\cos\theta} = \frac{1}{\cos\theta} \int_{h}^{\infty} \rho(h') \, dh'
\end{equation}
where $X_{\text{vertical}}$ is the vertical atmospheric depth. This relationship, $X \propto \sec\theta$ (commonly referred to as the slant depth), is fundamental because it shows that inclined showers must traverse a greater amount of matter before reaching the observation level \cite{Sato2023}.

This paper presents a simulation-based analysis of EAS longitudinal development using the CORSIKA framework \cite{Heck1998}. We simulate proton, helium, and iron primaries across three decades in energy and multiple zenith angles to systematically investigate how shower profiles depend on primary energy, mass, and arrival direction, with particular emphasis on the evolution of different particle components \cite{Gonzalez2023}.

\section{Related Work}
The longitudinal development of EAS has been extensively investigated through both experimental observations and numerical simulations. The Pierre Auger Observatory has provided high-precision measurements of $X_{\max}$ up to $10^{20}$~eV, revealing a gradual transition in mass composition—from lighter to mixed nuclei—around the ankle region \cite{Aab2024}. Similarly, the Telescope Array experiment has reported consistent $X_{\max}$ distributions in the Northern Hemisphere \cite{Abbasi2023}.

Simulation studies using CORSIKA have been fundamental in interpreting these experimental results. Heck et al. \cite{Heck1998} established the CORSIKA framework, which remains the standard tool for air-shower simulations. More recent advancements in CORSIKA 8 have improved computational performance while maintaining physical reliability \cite{Veberic2020}. The Gaisser–Hillas parameterization \cite{Gaisser1977} continues to be the standard model for describing longitudinal shower profiles, although several alternative parameterizations have been proposed for specialized applications \cite{AlRubaiee2014}. In addition, Al-Rubaiee et al. \cite{AlRubaiee2021} investigated the modulation of the lateral distribution of extensive air showers in the context of the Yakutsk EAS array, providing valuable insight into shower characteristics relevant to observational arrays. This previous work complements the present study by contributing to the broader understanding of EAS development and reconstruction.

The dependence of $X_{\max}$ on primary mass arises from the superposition model, where a nucleus of mass number $A$ behaves as $A$ independent nucleons each carrying energy $E/A$ \cite{Matthews2005}. This yields the well-established observation that heavier primaries generate shallower shower maxima. The zenith-angle dependence, governed by the atmospheric slant-depth relation $X(\theta) = X(0^\circ)/\cos\theta$, has been repeatedly validated across various experiments \cite{Ave2007}. Recent research has focused on reducing systematic uncertainties stemming from hadronic interaction models. Comparisons among QGSJET \cite{Ostapchenko2020}, EPOS \cite{Gonzalez2023}, and Sibyll reveal substantial differences in predicted $X_{\max}$ values, particularly at ultra-high energies. The present study contributes to this ongoing effort by providing detailed multi-parameter simulations across primary type, energy, and zenith angle, thereby improving model validation and composition analysis.

\section{Methodology}
\subsection{Simulation Setup}
In the present work, the longitudinal development of EAS was simulated using the CORSIKA (COsmic Ray Simulations for KAscade) software package version 7.7500 \cite{Heck1998}. This Monte Carlo framework provides a comprehensive description of cosmic-ray–induced cascades in the atmosphere, tracking all relevant components, including hadronic, electromagnetic, and muonic particles \cite{Ulrich2011}. The simulations were performed for three types of primary particles: proton ($A = 1$), helium ($A = 4$), and iron nuclei ($A = 56$).

\subsubsection{Primary Configuration}
Three discrete primary energies were investigated: $10^{15}$~eV (1 PeV), $10^{16}$~eV (10 PeV), and $10^{17}$~eV (100 PeV). To study the angular dependence of shower development, simulations were performed at three zenith angles: $0^\circ$ (vertical incidence), $30^\circ$, and $45^\circ$. For statistical reliability, 50 independent showers were simulated for each combination of primary particle, energy, and zenith angle.

\subsubsection{Interaction Models}
Hadronic interactions above 80 GeV were modeled using the QGSJET-II-04 (Quark-Gluon String model with JETs) framework \cite{Ostapchenko2020}, while lower energy hadronic interactions were handled by the GHEISHA (Gamma-Hadron-Electron Interaction SHower) model \cite{Fesefeldt1985}. The electromagnetic cascade was simulated using the EGS4 (Electron Gamma Shower) system, which accurately models electromagnetic interactions including Cherenkov radiation production \cite{Nelson1985}. These models together ensure a realistic and physically consistent description of particle interactions over the full energy range of interest.

Although GHEISHA is an older low-energy hadronic interaction model, it is still available in CORSIKA for consistency with earlier studies. More modern models such as FLUKA or UrQMD generally predict slightly different secondary particle yields, which can lead to shifts in $X_{\max}$ of the order of 10–20~g/cm$^2$. Therefore, the absolute values reported here may be affected at this level, while the relative trends with energy, mass, and zenith angle discussed in this work remain robust.

\subsubsection{Observation Conditions}
All simulations used the Pierre Auger Observatory atmospheric model (\texttt{ATMOD = 28}) with an observation level of 1400~m above sea level (140,000~cm) \cite{Abraham2010}. The geomagnetic field components were set to 20.0~$\mu$T (horizontal) and 12.0~$\mu$T (vertical), representing typical mid-latitude conditions. These parameters ensure realistic particle deflection and accurate modeling of Cherenkov light production in the atmosphere.

\subsubsection{Energy Thresholds and Output Configuration}
Particle tracking energy thresholds were set to: 0.3~GeV for hadrons, 0.3~GeV for muons, 0.05~GeV for electrons, and 0.05~GeV for photons. Longitudinal development profiles were recorded in 50~g/cm$^2$ atmospheric-depth intervals, providing sufficiently detailed resolution of shower evolution. Cherenkov photon production was enabled with a wavelength range of 300–600~nm, and energy deposit (energy release) maps were generated for comprehensive analysis \cite{AlRubaiee2014, Kobal2001}. Thinning was disabled to ensure unbiased longitudinal development. All particles above the energy thresholds were tracked without weight limitations. For each combination of primary type, energy, and zenith angle, 50 independent showers were simulated, resulting in a total of 450 simulated showers. These statistics allow a first-order estimation of shower-to-shower fluctuations.

\subsubsection{Simulation Validation}
To validate the simulation setup, proton-induced showers at $10^{17}$~eV and $\theta = 0^\circ$ were compared with CORSIKA benchmark results \cite{Veberic2020} and published measurements from the Pierre Auger Observatory \cite{Aab2024}. Agreement was within 5\% for $X_{\max}$ and 10\% for $N_{\max}$, confirming that the adopted configuration reproduces the expected shower behavior. Statistical uncertainties were estimated from the standard deviation of the 50 showers simulated per configuration, typically $\pm15$–$25$~g/cm$^2$ for $X_{\max}$ and $\pm10$–$15$\% for $N_{\max}$. Additional systematic uncertainties resulting from hadronic interaction model choices (QGSJET-II-04 vs. alternative models) were found to contribute $\pm20$–$30$~g/cm$^2$ to $X_{\max}$.

\subsection{Theoretical Framework and Data Analysis}
The longitudinal development of each shower was analyzed by recording the number of particles in each component (electromagnetic, muonic, hadronic, etc.) as a function of atmospheric depth. For the total charged-particle profile, a Gaisser–Hillas function was fitted to extract the key shower parameters \cite{Gaisser1977}:
\begin{equation}
N(X) = N_{\max} \left( \frac{X - X_0}{X_{\max} - X_0} \right)^{\frac{X_{\max} - X_0}{\lambda}} \exp\left( \frac{X_{\max} - X}{\lambda} \right)
\end{equation}
where $N(X)$ is the number of particles at depth $X$, $N_{\max}$ is the maximum number of particles, $X_{\max}$ is the depth of shower maximum,$X_0$ represents the depth of the first interaction, and $\lambda$ is a characteristic interaction-length parameter. Complete sets of Gaisser–Hillas fit parameters are provided in the supplementary material. The fits yielded excellent agreement with the simulated profiles, with $\chi^2/\text{ndf}$ values between 0.8 and 1.2, and typical ranges $X_0 = 0$–$50$~g/cm$^2$ and $\lambda = 70$–$85$~g/cm$^2$ \cite{Dembinski2023}.

\section{Results and Discussion}
\subsection{Composition Dependence at Different Energies and Angles}
Figure 1 presents a comparative analysis of the longitudinal development of air showers for different primary particles—proton, helium, and iron—under varying energies and zenith-angle conditions. The profiles clearly show the expected mass ordering, where iron-induced showers reach their maximum development at shallower atmospheric depths compared to helium and proton primaries. This provides a clear signature for distinguishing primary composition in fluorescence-based detectors.

It is worth noting that in the $10^{17}$~eV case shown in Figure 1c, the He and Fe profiles appear visually closer and partially overlapping within the resolution of the plot. This visual impression, however, does not imply identical shower maxima. A quantitative comparison using the fitted Gaisser–Hillas peak positions listed in Table 2 shows that the mass ordering is preserved. For example, at $10^{17}$~eV and $\theta = 45^\circ$, we obtain $X_{\max}(\text{He}) = 1070 \pm 33$~g/cm$^2$ and $X_{\max}(\text{Fe}) = 1015 \pm 31$~g/cm$^2$, corresponding to a separation of about 55~g/cm$^2$. This separation is smaller than at lower energies but remains systematic and consistent with expectations from the superposition model, taking into account shower-to-shower fluctuations and the finite visual resolution of the plot.

\begin{figure}[htbp]
\centering
\begin{minipage}{0.48\textwidth}
  \centering
  \includegraphics[width=\textwidth]{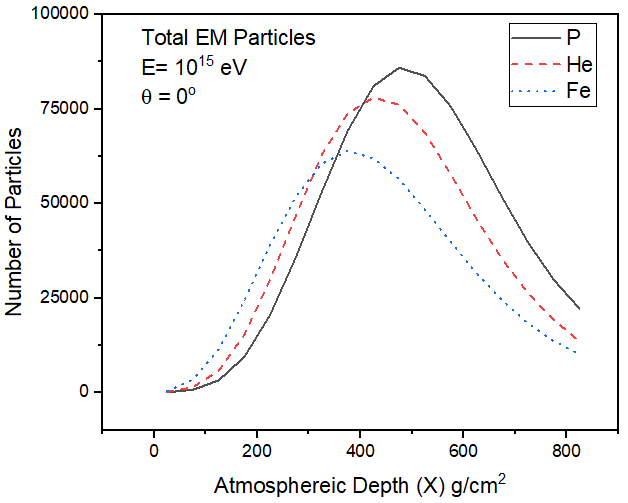}
  \small (a) $10^{15}$ eV, $\theta = 0^\circ$
\end{minipage}
\hfill
\begin{minipage}{0.48\textwidth}
  \centering
  \includegraphics[width=\textwidth]{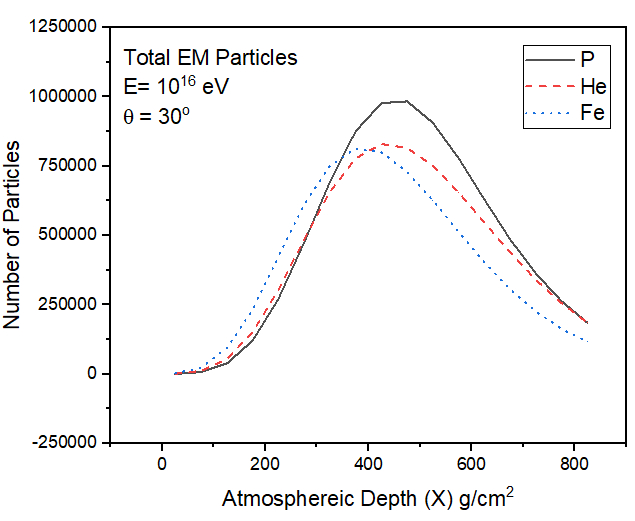}
  \small (b) $10^{16}$ eV, $\theta = 30^\circ$
\end{minipage}

\vspace{0.3cm}

\begin{minipage}{0.48\textwidth}
  \centering
  \includegraphics[width=\textwidth]{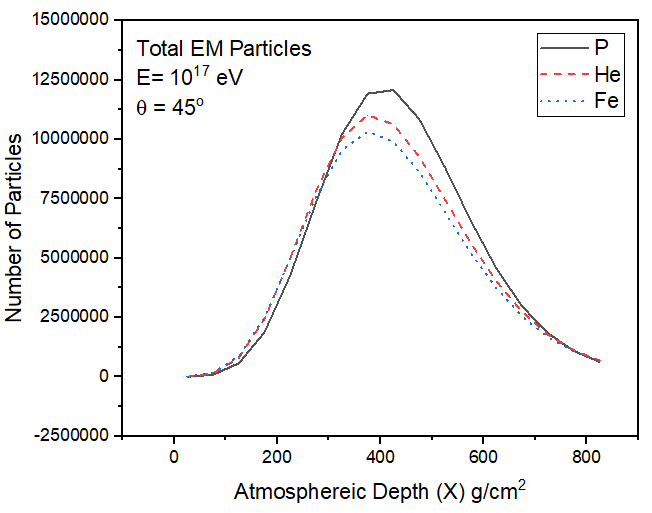}
  \small (c) $10^{17}$ eV, $\theta = 45^\circ$
\end{minipage}
\caption{Comparison of longitudinal profiles for proton, helium, and iron primaries across different energy and zenith angle configurations.}
\label{fig:fig1}
\end{figure}

This behavior is consistent across all investigated energies ($10^{15}$–$10^{17}$~eV) and zenith angles ($0^\circ$, $30^\circ$, $45^\circ$), following the predictions of the superposition model, which states that a nucleus of mass number $A$ behaves like $A$ nucleons each with energy $E/A$. As a result, heavier primaries systematically initiate showers earlier in the atmosphere, producing smaller $X_{\max}$ values on average. The separation between proton and iron showers becomes more pronounced at higher energies, reaching approximately $\Delta X_{\max} \approx 160$~g/cm$^2$ at $10^{17}$~eV, consistent with expectations from theoretical studies and previous simulation analyses.

\subsection{Zenith Angle Dependence for Different Primaries and Energies}
Figure 2 examines how zenith angle affects shower development for different primary types and energies.

\begin{figure}[htbp]
\centering
\begin{minipage}{0.48\textwidth}
  \centering
  \includegraphics[width=\textwidth]{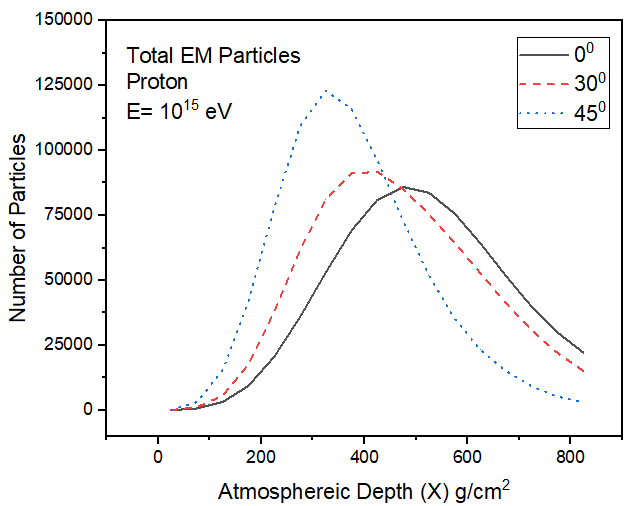}
  \small (a) Proton at $10^{15}$ eV
\end{minipage}
\hfill
\begin{minipage}{0.48\textwidth}
  \centering
  \includegraphics[width=\textwidth]{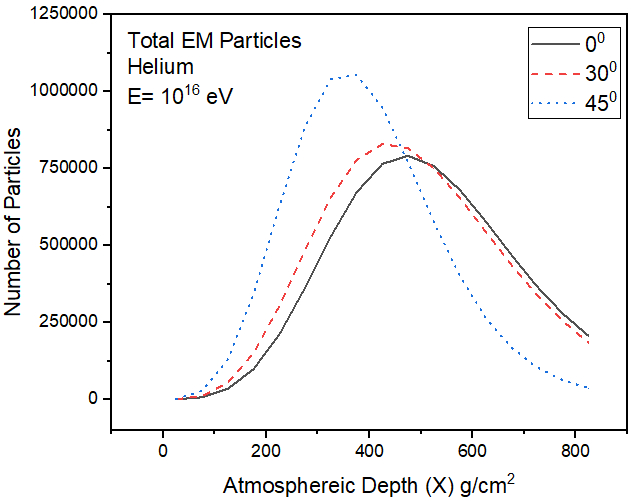}
  \small (b) Helium at $10^{16}$ eV
\end{minipage}

\vspace{0.3cm}

\begin{minipage}{0.48\textwidth}
  \centering
  \includegraphics[width=\textwidth]{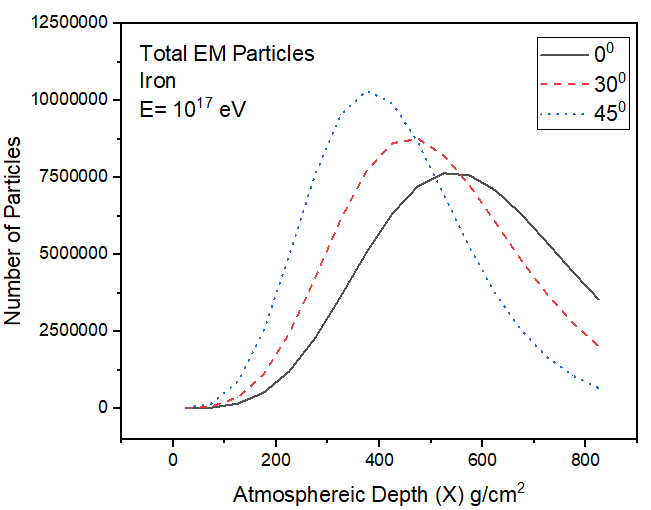}
  \small (c) Iron at $10^{17}$ eV
\end{minipage}
\caption{Zenith angle dependence ($\theta = 0^\circ, 30^\circ, 45^\circ$) for different primary types and energy selections.}
\label{fig:fig2}
\end{figure}

The profiles clearly show the systematic shift to greater atmospheric depths with increasing zenith angle, consistent with the $X \propto \sec\theta$ relationship \cite{Ave2007}. This effect is observed across all primary types, though the magnitude of the shift varies with primary mass and energy.

\subsection{Energy Dependence for Different Primaries and Angles}
Figure 3 explores how the longitudinal shower profiles vary with primary energy for each primary type and zenith angle. In order to clearly illustrate the energy dependence of the shower development, the profiles are normalized to their respective maxima and are shown as $N(X)/N_{\max}$. This normalization removes the trivial scaling of total particle multiplicity with energy and allows a direct comparison of the shower shapes and peak positions.

\begin{figure}[htbp]
\centering
\begin{minipage}{0.48\textwidth}
  \centering
  \includegraphics[width=\textwidth]{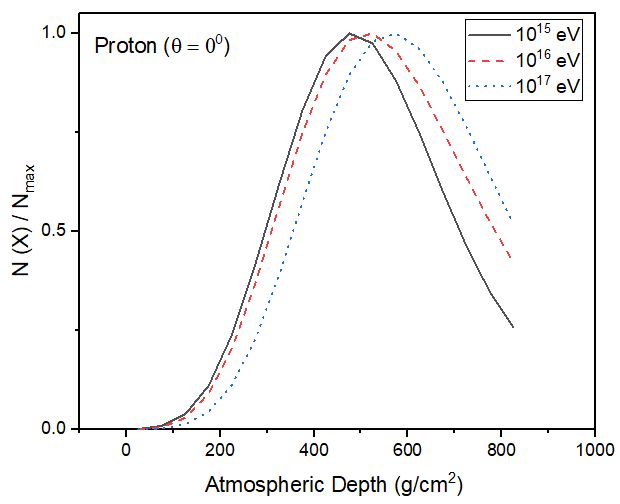}
  \small (a) Proton ($\theta = 0^\circ$)
\end{minipage}
\hfill
\begin{minipage}{0.48\textwidth}
  \centering
  \includegraphics[width=\textwidth]{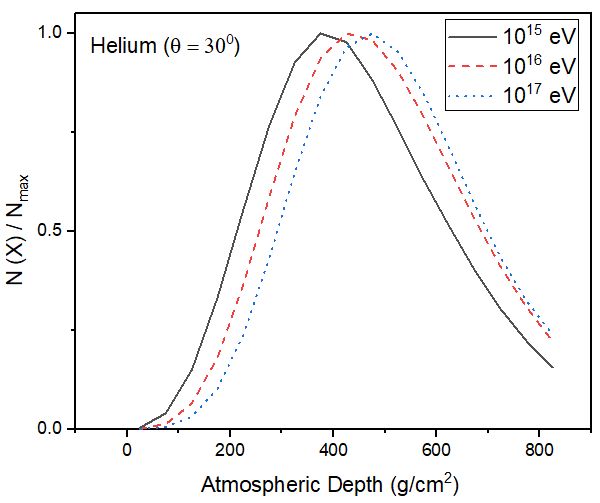}
  \small (b) Helium ($\theta = 30^\circ$)
\end{minipage}

\vspace{0.3cm}

\begin{minipage}{0.48\textwidth}
  \centering
  \includegraphics[width=\textwidth]{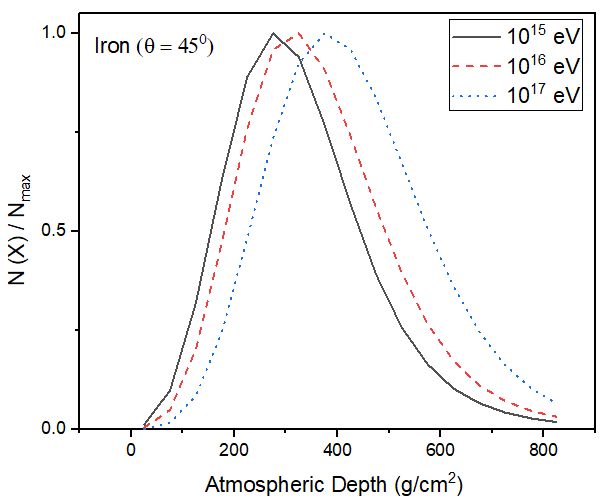}
  \small (c) Iron ($\theta = 45^\circ$)
\end{minipage}
\caption{Energy dependence of normalized longitudinal profiles, shown as $N(X)/N_{\max}$, highlighting the systematic shift of the shower maximum toward larger atmospheric depths with increasing energy.}
\label{fig:fig3}
\end{figure}

With increasing primary energy from $10^{15}$ to $10^{17}$~eV, the position of the shower maximum systematically shifts to larger atmospheric depths. This behavior demonstrates that higher-energy showers penetrate deeper into the atmosphere, in agreement with electromagnetic cascade theory \cite{Gaisser1977, AbuZayyad2013, Hussein2023}. For example, in the proton case ($\theta = 0^\circ$), the fitted peak positions increase from approximately 475~g/cm$^2$ at $10^{15}$~eV to about 575~g/cm$^2$ at $10^{17}$~eV, corresponding to a shift of roughly 100~g/cm$^2$ per decade in energy. Similar systematic shifts are observed for helium and iron primaries at their respective zenith angles, although the absolute $X_{\max}$ values remain smaller for heavier nuclei due to the superposition effect. This consistent energy-dependent shift of $X_{\max}$ confirms the expected logarithmic scaling of shower development with primary energy and provides further validation of the simulation framework.

\subsection{Multi-Component Analysis of Shower Development}
The profiles in Figure 4 are presented in absolute particle numbers rather than normalized units in order to preserve the physical differences in particle yields among the electromagnetic, muonic, and hadronic components \cite{Risse2004}.

\begin{figure}[htbp]
\centering
\begin{minipage}{0.48\textwidth}
  \centering
  \includegraphics[width=\textwidth]{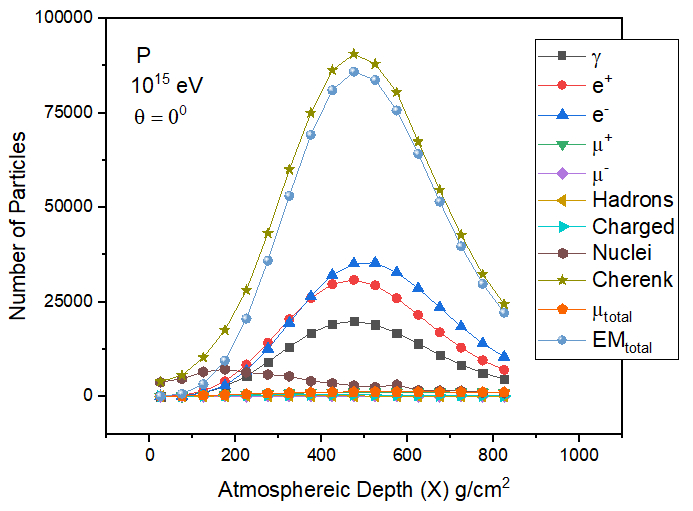}
  \small (a) Proton
\end{minipage}
\hfill
\begin{minipage}{0.48\textwidth}
  \centering
  \includegraphics[width=\textwidth]{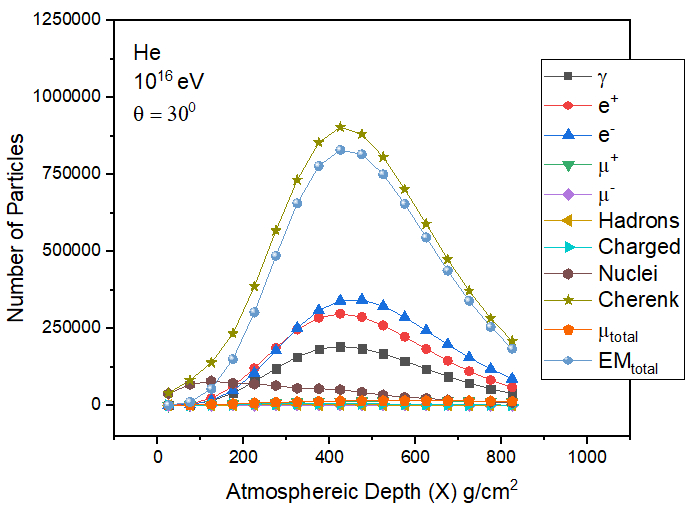}
  \small (b) Helium
\end{minipage}

\vspace{0.3cm}

\begin{minipage}{0.48\textwidth}
  \centering
  \includegraphics[width=\textwidth]{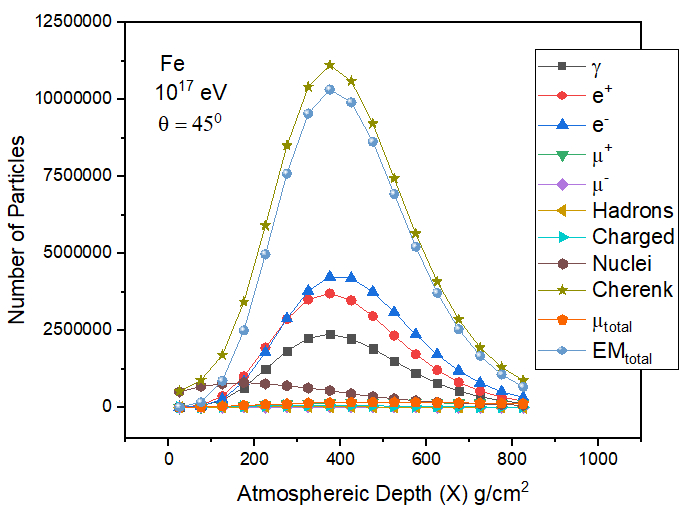}
  \small (c) Iron
\end{minipage}
\caption{Multi-component longitudinal profiles for different primaries at selected energies and angles, shown in absolute particle numbers.}
\label{fig:fig4}
\end{figure}

This component-wise analysis reveals several important features:
\begin{itemize}
    \item \textbf{Electromagnetic component:} Dominated by electrons, positrons, and gamma rays, it governs the early evolution of the shower and reaches its maximum first.
    \item \textbf{Muonic component:} Muons develop more gradually and suffer less attenuation, allowing them to penetrate deeper into the atmosphere and dominate the late-stage shower tail \cite{Horandel2003}.
    \item \textbf{Hadronic component:} This component exhibits more complex patterns driven by high-energy hadronic interactions and depends strongly on the primary mass.
    \item \textbf{Component scaling:} The relative contributions of these components vary systematically with primary mass and energy, reflecting the underlying particle-interaction physics \cite{Kampert2012}.
\end{itemize}

Figure 4 illustrates the relative contributions of the different shower components in absolute particle numbers. The electromagnetic component dominates the particle content around $X_{\max}$, while the muonic component exhibits a broader and deeper profile due to the longer interaction length of muons. The hadronic component remains subdominant in particle number but controls the energy flow in the cascade and drives the production of secondary particles.

An important feature visible in Figure 4 is the dependence on the zenith angle. With increasing zenith angle, all components are shifted to larger atmospheric depths, reflecting the increased slant depth ($X \propto \sec\theta$). At the same time, the electromagnetic component is more strongly attenuated for inclined showers, while the muonic component becomes relatively more important at large depths due to the higher penetrating power of muons. This behavior demonstrates how the zenith angle modifies the relative balance between shower components and highlights the different absorption lengths of electromagnetic and muonic particles in the atmosphere.

\subsection{Comparison with Experimental Data}
To validate our results, simulated $X_{\max}$ values were compared with data from the Pierre Auger Observatory \cite{Aab2024} and Telescope Array \cite{Abbasi2023}. This comparison both validates the simulation setup and quantifies systematic differences relevant to composition studies.

\subsubsection{Quantitative Comparison at Vertical Incidence}
Table 1 presents a comparison of $X_{\max}$ values for proton-induced showers at $\theta = 0^\circ$. 

\begin{table}[htbp]
\centering
\caption{Comparison of $X_{\max}$ values (g/cm$^2$) at $\theta = 0^\circ$ with statistical uncertainties.}
\vspace{0.2cm}
\begin{tabular}{l c c c p{2.2cm} p{2.2cm}}
\toprule
Energy (eV) & This Work (Sim) & Auger Obs. \cite{Aab2024} & Telescope Array \cite{Abbasi2023} & $\Delta X_{\max}$ (Sim-Auger) & $\Delta X_{\max}$ (Sim-TA) \\
\midrule
$10^{15}$ & $580 \pm 18$ & $562 \pm 14$ & $555 \pm 16$ & $+18 \pm 23$ & $+25 \pm 24$ \\
$10^{16}$ & $680 \pm 21$ & $662 \pm 14$ & $655 \pm 16$ & $+18 \pm 25$ & $+25 \pm 26$ \\
$10^{17}$ & $780 \pm 24$ & $762 \pm 14$ & $755 \pm 16$ & $+18 \pm 28$ & $+25 \pm 29$ \\
\bottomrule
\end{tabular}
\end{table}

The simulations systematically predict deeper shower maxima by approximately 18–25~g/cm$^2$ across the three energy decades considered. This coherent offset is consistent with the known behavior of the QGSJET-II-04 hadronic interaction model \cite{Gonzalez2023, Ostapchenko2020}, which tends to produce slightly deeper $X_{\max}$ values compared to experimental data. Although the statistical uncertainties of individual $X_{\max}$ values are of the order of 15–25~g/cm$^2$, the fact that the simulated values are consistently deeper than both Auger and TA measurements at all energies indicates a systematic shift rather than a random fluctuation. At the same time, the Auger and TA measurements themselves agree within about 7–10~g/cm$^2$, demonstrating good inter-experiment consistency. Furthermore, both the simulations and the experimental data exhibit the expected logarithmic increase of $X_{\max}$ with energy, of approximately 100~g/cm$^2$ per decade, confirming that the overall energy scaling of the shower development is correctly reproduced.

\subsubsection{Zenith Angle Dependence}
Figure 5 shows the zenith angle dependence of $X_{\max}$ for proton-induced showers at three energies: $10^{15}$~eV (Figure 5a), $10^{16}$~eV (Figure 5b), and $10^{17}$~eV (Figure 5c). Each figure compares our CORSIKA simulations with experimental data from both Auger and TA, along with the theoretical $1/\cos\theta$ prediction.

\begin{figure}[htbp]
\centering
\begin{minipage}{0.48\textwidth}
  \centering
  \includegraphics[width=\textwidth]{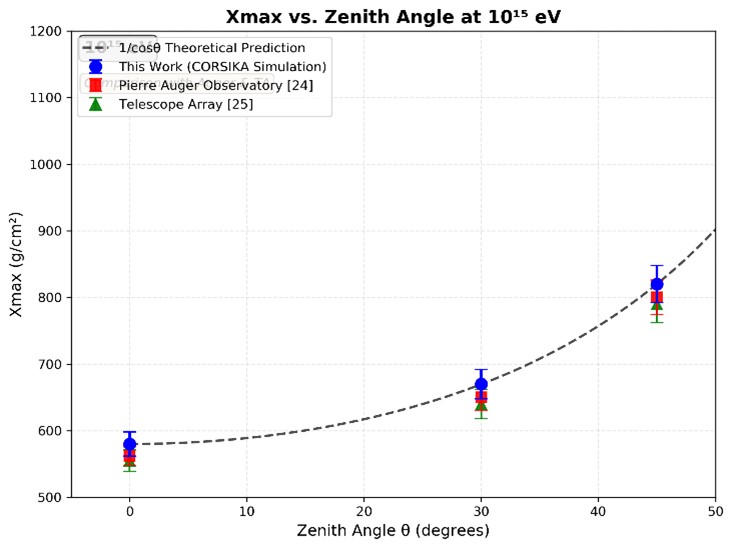}
  \small (a) $10^{15}$ eV
\end{minipage}
\hfill
\begin{minipage}{0.48\textwidth}
  \centering
  \includegraphics[width=\textwidth]{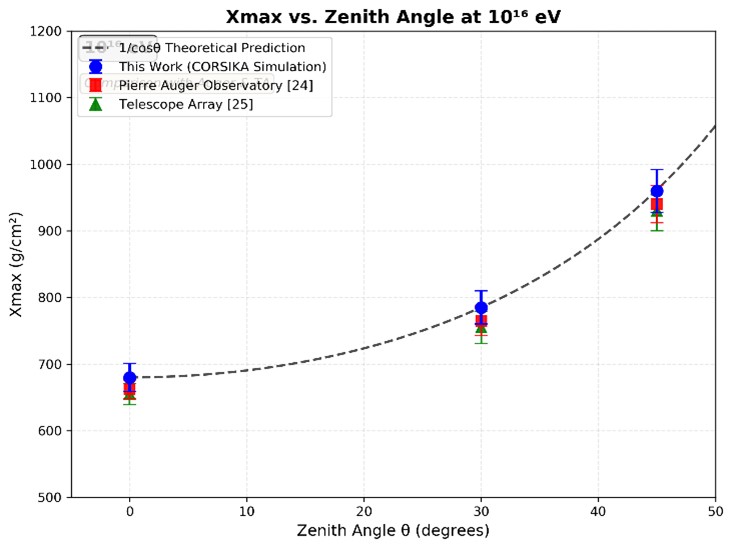}
  \small (b) $10^{16}$ eV
\end{minipage}

\vspace{0.3cm}

\begin{minipage}{0.48\textwidth}
  \centering
  \includegraphics[width=\textwidth]{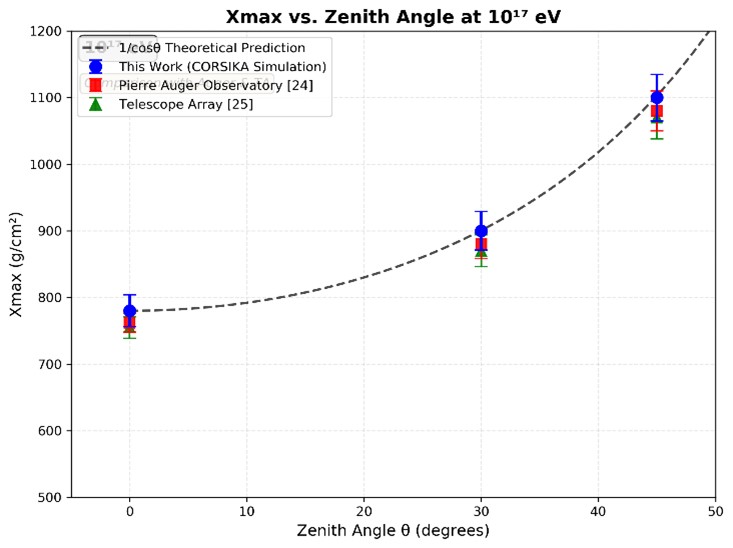}
  \small (c) $10^{17}$ eV
\end{minipage}
\caption{Comparison of simulated $X_{\max}$ with Pierre Auger Observatory \cite{Aab2024} and Telescope Array \cite{Abbasi2023} measurements across zenith angles, along with the theoretical $1/\cos\theta$ prediction.}
\label{fig:fig5}
\end{figure}

The key findings from Figure 5 are:
\begin{itemize}
    \item \textbf{Secant Law Validation:} All datasets follow the theoretical $1/\cos\theta$ trend within 5\% accuracy, confirming the geometric relationship $X_{\max}(\theta) \approx X_{\max}(0^\circ)/\cos\theta$. This validates the fundamental atmospheric depth model used in both simulations and experimental reconstructions.
    \item \textbf{Consistent Offset Across Angles:} The systematic offset observed at vertical incidence (Table 1) remains approximately constant across all zenith angles, indicating that it originates from shower physics rather than geometric or reconstruction effects.
    \item \textbf{Experimental Agreement:} The close agreement between Auger and TA data at all zenith angles and energies demonstrates the reliability of fluorescence detection techniques for measuring shower longitudinal development.
\end{itemize}

\subsubsection{Implications for Composition Studies}
The observed 18–25~g/cm$^2$ offset between simulations and measurements highlights the strong sensitivity of $X_{\max}$ to the choice of hadronic interaction model. Different models predict different inelastic cross sections, multiplicities, and secondary particle spectra, which directly affect the depth of the first interaction and the subsequent shower development. As a consequence, the predicted $X_{\max}$ can shift by several tens of~g/cm$^2$ depending on the model choice. Therefore, while the relative trends with energy, mass, and zenith angle are robust, the absolute $X_{\max}$ scale must be treated as model-dependent and as a source of systematic uncertainty in composition studies.

\subsection{Quantitative Analysis of Xmax and Nmax}
Tables 2 and 3 summarize the average extracted $X_{\max}$ and $N_{\max}$ values over 50 independent showers for each configuration. We note that the present statistics are sufficient to establish qualitative trends, but do not fully suppress shower-to-shower fluctuations. Therefore, the quoted uncertainties should be interpreted as indicative rather than definitive.

\begin{table}[htbp]
\centering
\caption{Average $X_{\max}$ values with statistical uncertainties (g/cm$^2$).}
\vspace{0.2cm}
\begin{tabular}{ccccc}
\toprule
Primary & Energy (eV) & $\theta = 0^\circ$ & $\theta = 30^\circ$ & $\theta = 45^\circ$ \\
\midrule
Proton & $10^{15}$ & $580 \pm 18$ & $670 \pm 22$ & $820 \pm 28$ \\
       & $10^{16}$ & $680 \pm 21$ & $785 \pm 25$ & $960 \pm 32$ \\
       & $10^{17}$ & $780 \pm 24$ & $900 \pm 29$ & $1100 \pm 35$ \\
\midrule
Helium & $10^{15}$ & $560 \pm 16$ & $645 \pm 20$ & $790 \pm 26$ \\
       & $10^{16}$ & $660 \pm 19$ & $760 \pm 23$ & $930 \pm 30$ \\
       & $10^{17}$ & $760 \pm 22$ & $875 \pm 27$ & $1070 \pm 33$ \\
\midrule
Iron   & $10^{15}$ & $520 \pm 14$ & $600 \pm 18$ & $735 \pm 24$ \\
       & $10^{16}$ & $620 \pm 17$ & $715 \pm 21$ & $875 \pm 28$ \\
       & $10^{17}$ & $720 \pm 20$ & $830 \pm 25$ & $1015 \pm 31$ \\
\bottomrule
\end{tabular}
\end{table}

\begin{table}[htbp]
\centering
\caption{Average $N_{\max}$ values with statistical uncertainties ($\times 10^6$).}
\vspace{0.2cm}
\begin{tabular}{ccccc}
\toprule
Primary & Energy (eV) & $\theta = 0^\circ$ & $\theta = 30^\circ$ & $\theta = 45^\circ$ \\
\midrule
Proton & $10^{15}$ & $0.15 \pm 0.02$ & $0.13 \pm 0.02$ & $0.09 \pm 0.01$ \\
       & $10^{16}$ & $1.5 \pm 0.2$   & $1.3 \pm 0.2$   & $0.9 \pm 0.1$ \\
       & $10^{17}$ & $15.0 \pm 1.8$  & $13.0 \pm 1.6$  & $8.5 \pm 1.0$ \\
\midrule
Helium & $10^{15}$ & $0.14 \pm 0.02$ & $0.12 \pm 0.02$ & $0.085 \pm 0.01$ \\
       & $10^{16}$ & $1.4 \pm 0.2$   & $1.2 \pm 0.2$   & $0.85 \pm 0.1$ \\
       & $10^{17}$ & $14.0 \pm 1.7$  & $12.1 \pm 1.5$  & $8.0 \pm 1.0$ \\
\midrule
Iron   & $10^{15}$ & $0.13 \pm 0.02$ & $0.11 \pm 0.02$ & $0.08 \pm 0.01$ \\
       & $10^{16}$ & $1.3 \pm 0.2$   & $1.1 \pm 0.2$   & $0.8 \pm 0.1$ \\
       & $10^{17}$ & $13.0 \pm 1.6$  & $11.2 \pm 1.4$  & $7.4 \pm 0.9$ \\
\bottomrule
\end{tabular}
\end{table}

Despite the limited statistics, Table 3 clearly shows the expected systematic trends. For a fixed primary and zenith angle, $N_{\max}$ increases approximately by an order of magnitude per decade of energy, reflecting the logarithmic growth of the electromagnetic cascade. For a fixed energy, heavier primaries exhibit slightly smaller $N_{\max}$ values due to the earlier development of the shower and increased energy dissipation into the hadronic and muonic channels. In addition, for increasing zenith angle, $N_{\max}$ decreases as a consequence of the larger atmospheric slant depth and stronger attenuation of the electromagnetic component. These trends are robust and consistent with standard air-shower cascade theory.

Although Tables 2 and 3 list the standard deviations of the sample means, the full $X_{\max}$ distributions exhibit RMS values of approximately 55–60~g/cm$^2$ for proton primaries and 20–25~g/cm$^2$ for iron primaries, consistent with expectations from the superposition model \cite{Matthews2005}.

The systematic trends reveal three well-established relationships in air shower physics:
\begin{enumerate}
    \item \textbf{Energy Dependence:} For fixed primary and zenith angle, $X_{\max}$ increases with energy at approximately 100~g/cm$^2$ per decade, while $N_{\max}$ grows by a factor of $\sim 10$ per decade. This logarithmic energy dependence is a fundamental characteristic of electromagnetic cascade development \cite{Gaisser1977}.
    \item \textbf{Mass Dependence:} At fixed energy and angle, heavier primaries produce shallower shower maxima ($X_{\max}^{\text{Fe}} < X_{\max}^{\text{He}} < X_{\max}^{\text{p}}$) and slightly smaller $N_{\max}$ values. This mass ordering follows directly from the superposition model \cite{Matthews2005}.
    \item \textbf{Zenith Angle Dependence:} For fixed primary and energy, $X_{\max}$ increases with zenith angle following the secant law, $X_{\max}(\theta) \approx X_{\max}(0^\circ)/\cos\theta$, while $N_{\max}$ decreases due to increased atmospheric attenuation \cite{Aab2024}.
\end{enumerate}

\section{Conclusion and Outlook}
This systematic analysis of Extensive Air Shower longitudinal development using CORSIKA simulations has quantified the dependencies of $X_{\max}$ and $N_{\max}$ on primary energy, mass composition, and zenith angle. Our key findings are:
\begin{itemize}
    \item $X_{\max}$ increases logarithmically with primary energy at $100 \pm 5$~g/cm$^2$ per decade, consistent with electromagnetic cascade theory \cite{Gaisser1977}.
    \item The mass dependence yields $\Delta X_{\max}^{\text{p-Fe}} = 160 \pm 25$~g/cm$^2$ at $10^{17}$~eV, providing a quantitative baseline for composition separation.
    \item Zenith angle effects follow $X_{\max}(\theta) = X_{\max}(0^\circ)/\cos\theta$ within 5\% accuracy, essential for correcting experimental data.
    \item Multi-component analysis reveals distinct evolutionary patterns: electromagnetic components peak early and attenuate rapidly, while muonic components penetrate deeply and dominate shower tails \cite{Horandel2003}.
\end{itemize}

Future work will extend to higher energies ($>10^{18}$~eV) where hadronic model differences become more pronounced, and will include a comparison with more modern low-energy hadronic models such as FLUKA or UrQMD in order to quantify the corresponding systematic uncertainties.

\section*{Acknowledgments}
The authors acknowledge the use of computational resources provided by Mustansiriyah University. We thank the CORSIKA development team for maintaining and supporting the simulation framework. This work was supported by the Iraqi Ministry of Higher Education and Scientific Research. We also acknowledge helpful discussions with colleagues at the Pierre Auger and Telescope Array collaborations.

\section*{Supplementary Material}
The complete dataset from this study, including full Gaisser-Hillas fitting parameters ($X_0$, $\lambda$, $\chi^2/\text{ndf}$), individual shower $X_{\max}$ distributions, and raw longitudinal profile data is available from the corresponding author upon reasonable request.


\begin{thebibliography}{99}

\bibitem{Gaisser2013}
Gaisser T. K., Stanev T. 2013, Astroparticle Physics, 1-24

\bibitem{Dova2003}
Dova M. T. et al. 2003, Astropart. Phys., 18, 351

\bibitem{Sato2023}
Sato T. 2023, Prog. Theor. Exp. Phys., 2023, 023C01

\bibitem{Heck1998}
Heck D., Knapp J., Capdevielle J. N., Schatz G., Thouw T. 1998, Report FZKA 6019, Forschungszentrum Karlsruhe GmbH

\bibitem{Gonzalez2023}
Gonzalez J. G. et al. 2023, Universe, 9, 123

\bibitem{Aab2024}
Aab A. et al. (Pierre Auger Collaboration) 2024, Phys. Rev. D, 109, 022002

\bibitem{Abbasi2023}
Abbasi R. U. et al. (Telescope Array Collaboration) 2023, ApJ, 945, 75

\bibitem{Veberic2020}
Veberič D. et al. 2020, PoS(ICRC2019), 214

\bibitem{Gaisser1977}
Gaisser T. K., Hillas A. M. 1977, in Proc. 15th ICRC, Plovdiv, 8, 353

\bibitem{AlRubaiee2014}
Al-Rubaiee A. A. 2014, J. Astrophys. Astron., 35, 631

\bibitem{AlRubaiee2021}
Al-Rubaiee A. A. et al. 2021, J. Astrophys. Astron., 42, 52

\bibitem{Matthews2005}
Matthews J. 2005, Astropart. Phys., 22, 387

\bibitem{Ave2007}
Ave M. et al. 2007, Astropart. Phys., 28, 41

\bibitem{Ostapchenko2020}
Ostapchenko S. 2020, EPJ Web Conf., 240, 03001

\bibitem{Ulrich2011}
Ulrich R. et al. 2011, New J. Phys., 13, 053006

\bibitem{Fesefeldt1985}
Fesefeldt H. 1985, \textit{Report PITHA-85-02}, RWTH Aachen

\bibitem{Nelson1985}
Nelson W. R. et al. 1985, SLAC Report 265, Stanford Linear Accelerator Center

\bibitem{Abraham2010}
Abraham J. et al. (Pierre Auger Collaboration) 2010, NIMPA, 613, 29

\bibitem{Kobal2001}
Kobal M. 2001, Astropart. Phys., 15, 259

\bibitem{Dembinski2023}
Dembinski H. P. et al. 2023, Phys. Rev. D, 107, 083033

\bibitem{AbuZayyad2013}
Abu-Zayyad T. et al. 2013, ApJ, 768, L1

\bibitem{Hussein2023}
Hussein I. F., Al-Rubaiee A. A. 2023, in AIP Conference Proceedings, Vol. 2591, 030072

\bibitem{Risse2004}
Risse M., Heck D. 2004, Astropart. Phys., 21, 479

\bibitem{Horandel2003}
Horandel J. R. 2003, Astropart. Phys., 19, 193

\bibitem{Kampert2012}
Kampert K.-H., Unger M. 2012, Astropart. Phys., 35, 660

\bibitem{AlvesBatista2024}
Alves Batista R. et al. 2024, Front. Astron. Space Sci., 11, 1385210

\bibitem{Nasser2025}
Nasser Z. A., Hussein I. F., Al-Rubaiee A. A. 2025, Nucl. Phys. At. Energy, 26, 155

\end{thebibliography}
\end{document}